%% file: isoline_arxiv.tex
\pdfoutput=1
\documentclass[11pt]{article}
\usepackage{amsmath}
\usepackage{graphicx}
\usepackage{natbib}
\usepackage{url}
\usepackage{color}

\begin{document}

\renewcommand{\mathbf} \vec

\markright{Computers \& Geosciences, 2009, \textbf{35}(10):2020-2031}

\title{Isoline Retrieval: An optimal sounding method for validation of advected contours}

\author{Peter Mills\\
Peteysoft Foundation\\
\textit{petey@peteysoft.org}
}

\maketitle

\begin{abstract}
\input{isoline_abstract.tex}

\end{abstract}


\input{isoline_body_els.tex}

\section*{Acknowledgements}

\input{isoline_ack.tex}

\appendix

\section{Proof of increased accuracy}
\label{apxA}
\input{iso_appendix1.tex}

\bibliography{isoline_ret}

\end{document}

%% file: isoline_abstract.tex
The study of chaotic mixing is important for its potential to improve
our understanding of fluid systems.  Contour advection simulations
provide a good model of the phenomenon by tracking the
evolution of one or more contours or isolines of a trace substance to a
high level of precision.  The most accurate method of validating an advected
contour is to divide the tracer concentration into discrete ranges and perform a
maximum likelihood classification, a method that we term, ``isoline 
retrieval.''  Conditional probabilities generated as a result provide
excellent error characterization.

In this study, a water vapour isoline of 0.001 mass-mixing-ratio is
advected over five days in the upper troposphere and compared with
high-resolution AMSU (Advanced Microwave Sounding Unit) satellite retrievals. 
The goal is to find the same fine-scale, chaotic mixing in the isoline
retrievals as seen in the advection simulations.
Some of the filaments generated by the simulations show
up in the conditional probabilities as areas of reduced probability.
By rescaling the probabilities, the filaments may be revealed in the isoline 
retrievals proper with little effect on the overall accuracy.
Limitations imposed by the specific context, i.e. water-vapour 
retrieved with AMSU in the upper troposphere, are discussed.
Nonetheless, isoline retrieval is shown to be a highly 
effective technique for atmospheric
sounding, showing good agreement with both 
ECMWF (European Centre for Medium-range Weather
Forecasts) assimilation data and radiosonde measurements.

{\color{red}Software for isoline retrieval can be found at: \url{http://isoret.sourceforge.net}}

%% file: isoline_body_els.tex
\section{Introduction}
Many
trace atmospheric constituents are inert enough that they may be modelled
as passive tracers--that is, there exist neither sources nor sinks.
Traditional methods for tracer advection track the concentration at fixed
points along a grid, the time evolution calculated by numerical integration
of partial differential equations.  The disadvantage of these 
Eulerian methods is that the horizontal resolution is often quite limited since
improvements are paid for to the third power in computational speed.

Lagrangian models track the concentration along moving air-parcels.  
Contour advection, for example, is a powerful technique
that models the evolution of one or more contours or isolines of a passive
tracer. \citep{Dritschel1988,Dritschel1989}   The method is adaptive in the
sense that new points are added to or removed from the evolving contour in
order to maintain the integrity of the curve;  thus, its horizontal
configuration may be predicted to a high degree of precision.

\begin{figure}
  \centering
    \includegraphics[angle=90, width=0.95\textwidth]{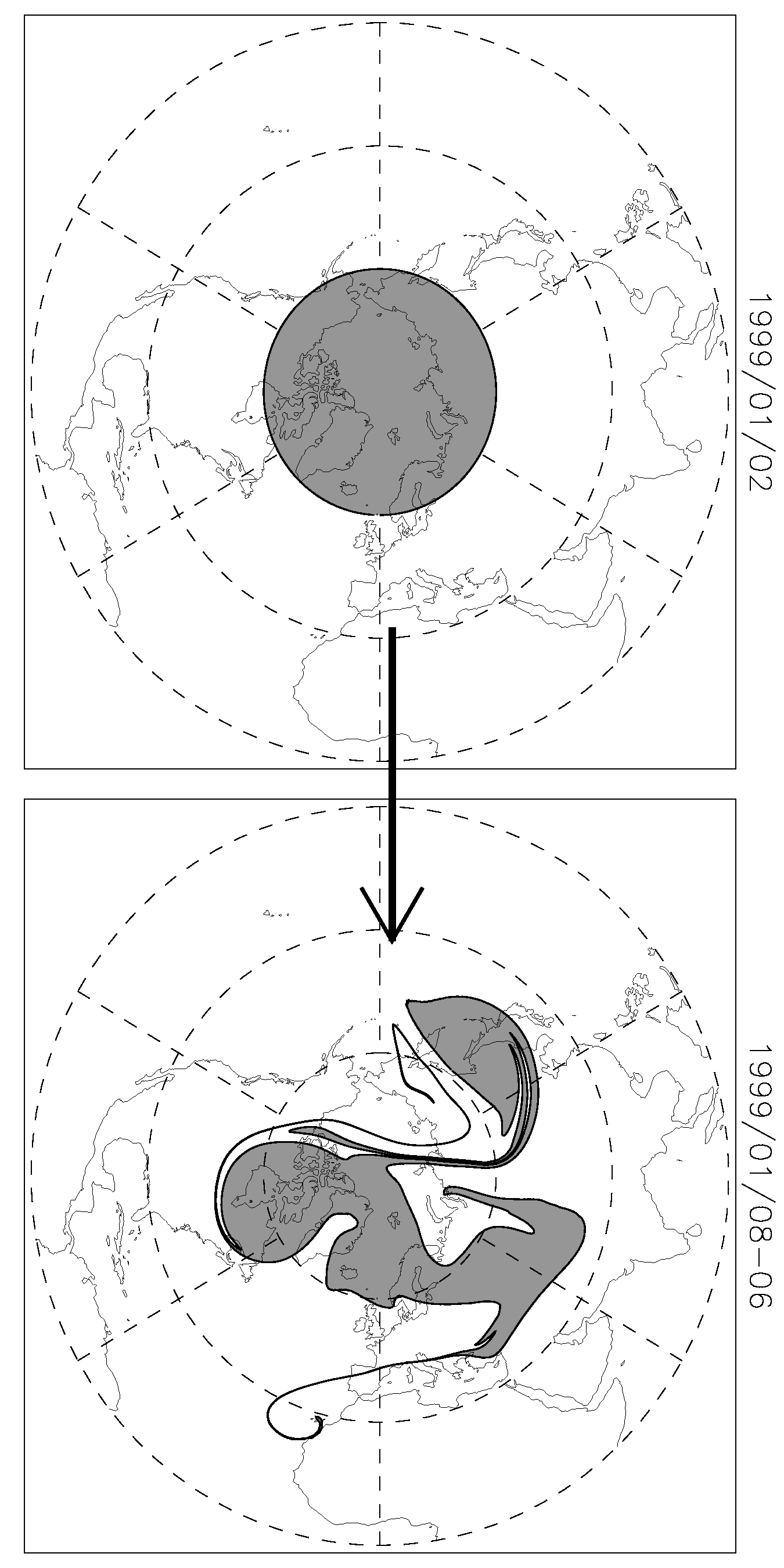}
  \caption{An example of contour advection}\label{contour_ex}
\end{figure}

Despite being driven by finitely resolved wind-fields, these advected
contours often show a great deal of fine-scale detail,
\citep{Waugh_Plumb1994,Methven_Hoskins1999} as shown
in Figure \ref{contour_ex}.  This results from a continual
process of stretching and folding, much like in a so-called ``baker's
map.'' \citep{Ott1993,Ottino1989} Many recent studies have attempted
to verify the existence of this fine-scale structure  
(necessarily ignored by most GCM's) 
in the real atmosphere, \citet{Appenzeller_etal1996}
being just one example.

Satellite instruments are one of the most prolific measurement sources for
trace atmospheric species.  They cannot measure the concentrations directly,
however;  rather they
measure the intensity of radiation in a narrow beam.
By understanding emission and absorption processes, the quantities 
of interest may be derived by performing some sort of inversion.  
In order to validate an advected contour, one should appreciate that
it is not necessary to know the exact value of the tracer at any given point.  
One needs only to determine whether the point falls within or without 
the contour--that is, is the concentration higher or lower 
than the value of the isoline?

This is a classification problem:  let $\mathbf{x}$ be a vector of measurements,
e.g. electro-magnetic radiances at several different frequencies.
The concentration (state variable in classical inverse theory \citep{Rodgers2000}
--here it is taken as a scalar) is divided into two or more broad ranges:
these will be distributed according to some conditional probability, $P(j |
\mathbf{x})$, where $j$ enumerates the range.  When doing retrievals, we seek
the most likely outcome, that is:
\begin{equation}
c=\arg \underset {j} {\max} P(j | \mathbf{x})
\label{class_mle}
\end{equation}
where c is the retrieved range of values. 
It is easy to show that a classification algorithm based on maximum
likelihood is the most accurate retrieval method possible for validating an
advected contour as demonstrated in Appendix A.  The conditional
probability, $P(j | \mathbf{x})$, is estimated by collecting
sample measurement vectors with corresponding concentrations which
have been converted to discrete values or \emph{classes}.  This is known as
the \emph{training data}.

The Advanced Microwave Sounding Unit (AMSU) series of satellite instruments 
 detect water vapour and oxygen at a high horizontal resolution.  
Because of the downward-looking measurement geometry, information on 
vertical variations is gained by measuring in several different
frequency bands or channels. The number of water-vapour channels is small:  
only five in the AMSU-B instrument implying a low vertical resolution, 
while contour advection must by necessity be limited to a single vertical
level.  On the other hand, the relatively few channels makes it ideal for
performing classification retrievals.

The purpose of this work is to perform contour advection simulations and compare
these with isolines of water vapour retrieved from the AMSU-A and -B satellite
instruments. While the region of sensitivity of these instruments (the
troposphere) means that the contour advection simulations will diverge after a
very short period, nonetheless it is hoped that the same fine-scale mixing seen
in the simulations will also show up in the retrievals.

\section{Data sources}

\subsection{The AMSU instrument}
\begin{table}
  \caption{NOAA AMSU channel frequencies.}\label{AMSU_channels}
  \centering
  \begin{tabular}[h]{|llll|}\hline
    channel & centre    & sideband & nominal \\
            & frequency & offset   & width
    \\ \hline
    AMSU-A: & [MHz] & [MHz] & [MHz] \\
    6 & 54400 & $\pm$ 105 & 190 \\
    7 & 54940 & $\pm$ 105 & 190 \\
    8 & 55500 & $\pm$ 87.5 & 155 \\
    9 & 57290 & $\pm$ 87.5 & 155 \\
    10 & 57290 & $\pm$ 217 & 77 \\
    AMSU-B: & [GHz] & [GHz] & [GHz] \\
    16 & 89.0 & $\pm$ 0.9 & 1.0 \\
    17 & 159.0 & $\pm$ 0.9 & 1.0 \\
    18 & 183.31 & $\pm$ 1.0 & 0.5 \\
    19 & 183.31 & $\pm$ 3.0 & 1.0 \\
    20 & 183.31 & $\pm$ 7.0 & 2.0 \\ \hline
    channel & centre    & sideband & nominal \\
            & frequency & offset   & width \\ \hline
  \end{tabular}
\end{table}
AMSU-B is a downward-looking instrument that detects microwave radiation
in five double side-band channels in the sub-millimeter range, all
sensitive to water-vapour.  Three are centred on the water-vapour emission line
at 183.31 GHz and two are surface looking, so-called ``window'' channels, also
sensitive to water-vapour and located at 89 and 150 GHz.  Since the instrument
is a sister to the AMSU-A instrument, the channels are typically numbered
between 16 and 20, starting with the two surface looking channels and then
the three 183 gigahertz channels, ending with the deepest-looking of the
three. \citep{Saunders_etal1995} Details of all five AMSU-B channels are
shown in Table \ref{AMSU_channels} along with a selected sub-set of the
AMSU-A channels. \citep{Rosenkranz2001,Kramer2002} These latter are
sensitive to oxygen thus supplying temperature information which is needed
for our retrievals.  See the Section \ref{rad_tran} on radiative transfer modelling.

The NOAA series of satellites upon which AMSU
is mounted fly in an 833 km sun-synchronous orbit with roughly 14 orbits
per day while the instrument uses a cross-track scanning geometry.  For AMSU-B,
there are 90 different viewing angles between -50 and 50 degrees,
sampling at a rate of roughly 22.5 scans per minute.  
This translates to over 300 000 measurements per
satellite per day at resolutions of between 15 and 50 km, 
while full global coverage requires only a single day of
data from three satellites.  The AMSU-A instrument has one ninth the
resolution as it has only 30 different viewing angles and scans at
one-third the rate.

\subsection{ECMWF data}

The European Centre for Medium-range Weather Forecasting (ECMWF) supplies a
synthesized, gridded data set based on an amalgam of in-situ, sonde,  
satellite and other remote-sensing measurements that have been fitted to
a dynamical model, hence the term ``assimilation'' data.
Gridding of the data used in this study is 1.5 by 1.5 degrees
longitude and latitude while it is laid out vertically along sixty so-called
``sigma'' levels. The idea behind sigma levels is that they are terrain
following close to the surface but revert to pressure levels at higher
altitudes.  The pressure at a given ECMWF sigma level is calculated as follows:
\begin{equation}
p_i=a_i + b_i p_0
\end{equation}
where $i$ is the level, $a_i$ and $b_i$ are constants and $p_0$
is the surface pressure.
For the most comprehensive and up-to-date
information on this product, please refer to the ECMWF website:
\url{http://www.ecmwf.int/research/ifsdocs/index.html}.

Included in the ECMWF data are gridded fields of temperature, pressure,
humidity and cloud content, all derived from the ERA-40 product.
These will be used both to validate retrievals and to generate the
training dataset.  For the latter, 
AMSU radiances are simulated with a radiative transfer model
and paired with water-vapour mixing-ratios from the input profiles. 
The zonal and meridional wind fields will drive the contour advection
simulation.

\subsection{Radiative transfer modelling}
\label{rad_tran}

Simulated training data is used instead of
actual AMSU measurements co-located with radiosonde measurements
for several reasons.  First, the irregular coverage of radiosonde
launch locations mean that statistics will not be representative.
Second, co-locations are never exact in time and space, especially
if balloon drift is not accounted for.  Moreover, satellite instruments
sample a large area while radiosondes will produce point measurements.
All these factors will make strictly empirical training data
quite noisy, while radiative transfer models have reached a
level of maturity that they can accurately model emissions 
if the atmospheric state is sufficiently well known.
In the case of training data, 
the atmospheric state need not be known exactly anyway;
it only has to represent a state that could reasonably occur
in the real atmosphere.

For a given vertical profile, the density of radiation at a single
frequency emitted to space in a single direction may be modelled via the
radiative transfer (RT) equation:
\begin{equation}
\frac{dI}{ds} = ( B - I) \sum_i \alpha_i \rho_i + \sigma \rho_s \left [
\int I(s, ~ \theta^\prime) P(\theta^\prime, ~ \theta) d \theta^\prime - I
\right ]
\label{rt_eq}
\end{equation}
where $I(\theta, ~s)$ is the intensity of radiation per solid angle, per unit frequency,
$s$ is the path (a function of altitude), $\alpha_i$ is the absorption
cross-section of the $i$th species, $\rho_i$ is the density of the $i$th species
and $B$ is the Planck function.  Equation
\ref{rt_eq} assumes a horizontally isotropic, scattering atmosphere 
so that $P$ is the scattering phase
function, which predicts the rate of transfer of radiation from the incoming
direction, $\theta^\prime$, to the outgoing direction, $\theta$, and $\sigma$ is
the scattering cross-section.
The density of the scatterer (here we assume there is
only one) is denoted by $\rho_s$.

Radiative transfer simulations were performed using RTTOV version 8, 
a fast radiative transfer simulator.  
The efficiency of RTTOV is gained by computing a 
linearized version of the RT equation for emission and absorption
while scattering by hydrometeors is treated by the Eddington approximation.
 \citep{Saunders_etal2005}

\section{Methods}
\subsection{Contour advection}

The evolution of a tracer in Lagrangian coordinates is given very
simply as:
\begin{eqnarray}
\frac{dq}{dt} & = & S(\mathbf{r}, t) \label{tracer}
\end{eqnarray}
where $q$ is the concentration of the tracer for a given parcel of air, 
$t$ is time and $S$ is the source term.  
The coordinate $\mathbf{r}$ evolves according
to the transport equation:
\begin{equation}
\frac{d \mathbf{r}}{dt} = \mathbf{v}(\mathbf{r}, t) \label{transport}
\end{equation}
where $\mathbf{v}$ is the velocity of the fluid.  Equation (\ref{tracer})
assumes that either the fluid is incompressible or the tracer concentration is
measured as a mixing-ratio.  In what is known as a conserved
or passive tracer, there are no source terms and the right-hand-side 
of (\ref{tracer}) will be zero.

In contour advection, only a single isoline is modelled at a time, where:
\begin{equation}
q \left [\mathbf{f}(s) \right ] = q_0
\end{equation}
defines the contour: $s$ is the path, $q_0$ is the value of the isoline and
$\mathbf{f}$ is the contour which evolves according to (\ref{transport}).  One
way to imagine it is to think of a blob of dye injected into a moving
fluid.  To first order, its evolution may be modelled by considering only
its outlines.  

It stands to reason that the function, $\mathbf{f}$, 
cannot be represented exactly; naturally, it can be defined to
arbitrary precision using a set of discrete points.
These are advected by integrating Equation
(\ref{transport}), typically with a Runge-Kutta scheme using velocities
interpolated from a grid. To maintain a constant precision, 
new points are added or removed at regular time intervals on the basis of 
some criterion or metric, either simple distance or curvature as in
\citet{Dritschel1988}.

We would like each pair of adjacent points to trace out a certain fraction of
arc (say, $\sim 1^\circ$) so that the curvature criterion may 
be defined as follows:
\begin{equation}
\alpha_{min} \le \frac{\Delta s}{r_c} \le \alpha_{max}
\label{new_point_crit}
\end{equation}
where $\alpha_{min}$ and $\alpha_{max}$ are the minimum and maximum allowed
fractions of arc respectively, $r_c$ is the radius of curvature and $\Delta s$
is the path difference between two adjacent points.  The number of new points
added will be in proportion to the ratio of the measure to the maximum, 
($=\Delta s /(r_c\alpha_{max})$) interpolated at regular intervals along the path.

Parametric fitting of $\mathbf{f}$ with a cubic spline \citep{nr_inc2} will
return a set of second order derivatives with respect to the path.  
These are then used
to calculate the curvature so that testing for new points and
interpolating them may be done in a single step.

\subsection{Adaptive Gaussian filtering}

\input{agf_simple_els.tex}

\subsection{Isoline retrieval}

\begin{figure}
  \centering
  \includegraphics[angle=90, width=0.95\textwidth]{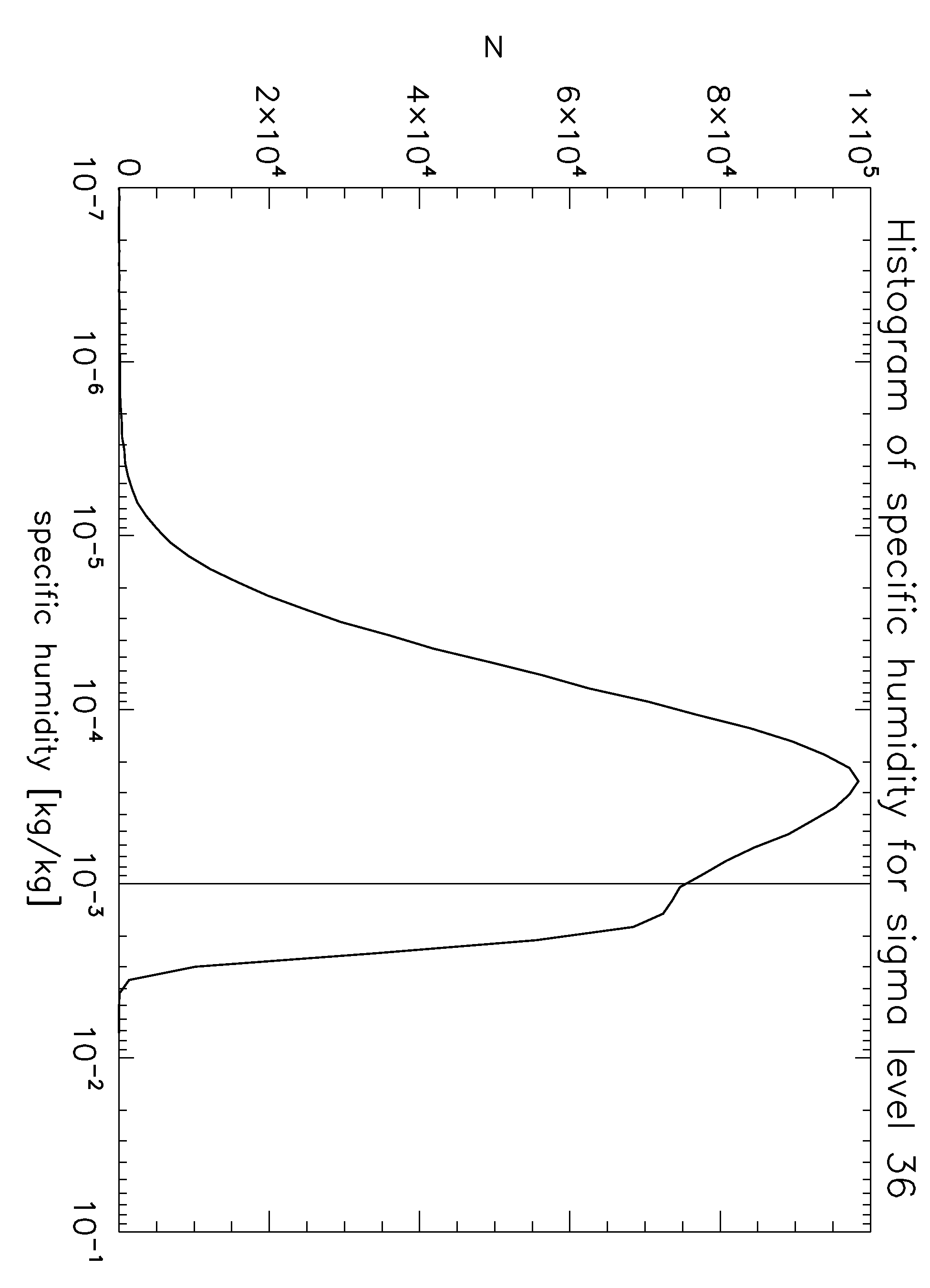}
  \caption{Histogram of ECMWF water-vapour mixing-ratios for sigma level 36.}\label{wv_vmr_hist}
\end{figure}
\begin{figure}
  \centering
  \includegraphics[angle=90, width=1\textwidth]{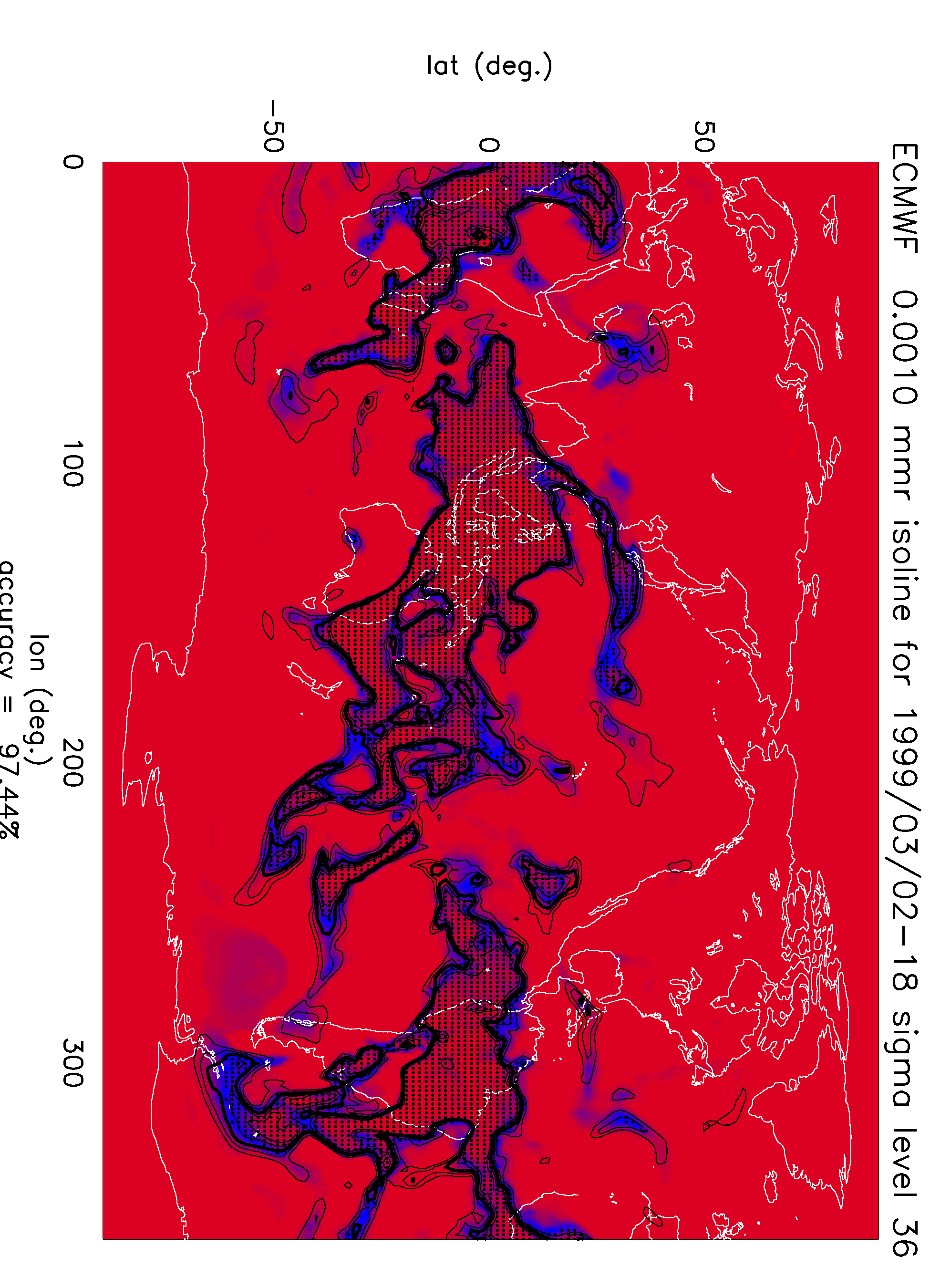}
  \includegraphics[angle=90, width=0.7\textwidth]{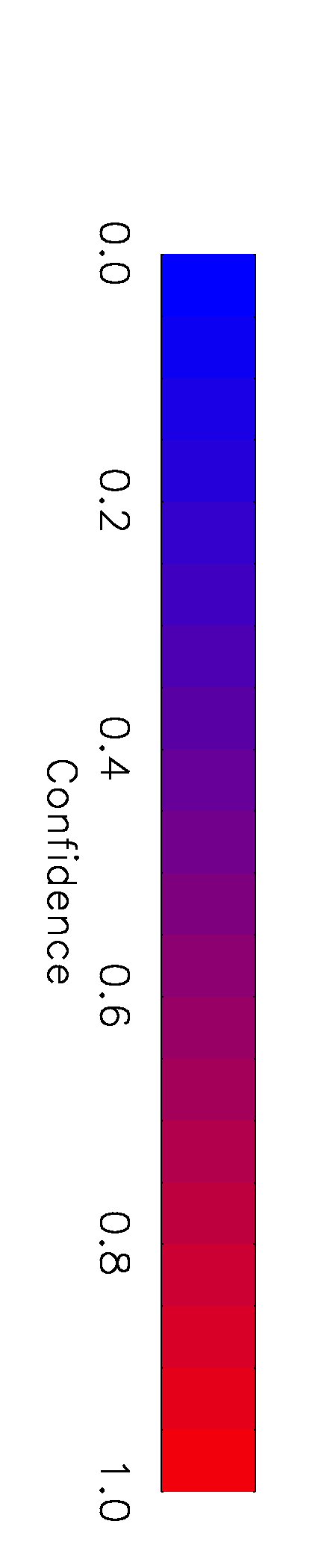}
  \caption{Test of isoline retrieval algorithm using radiances simulated from ECMWF data.  
Heavy black line is true isoline with auxiliary fine contours 
for 0.6, 0.8, 1.2 and 1.4 specific humidity levels.  
A dot indicates a positive ($c=2$) classification.  $W=30$}\label{sample_ret1}
\end{figure}
\begin{figure}
  \centering
  \includegraphics[width=0.95\textwidth]{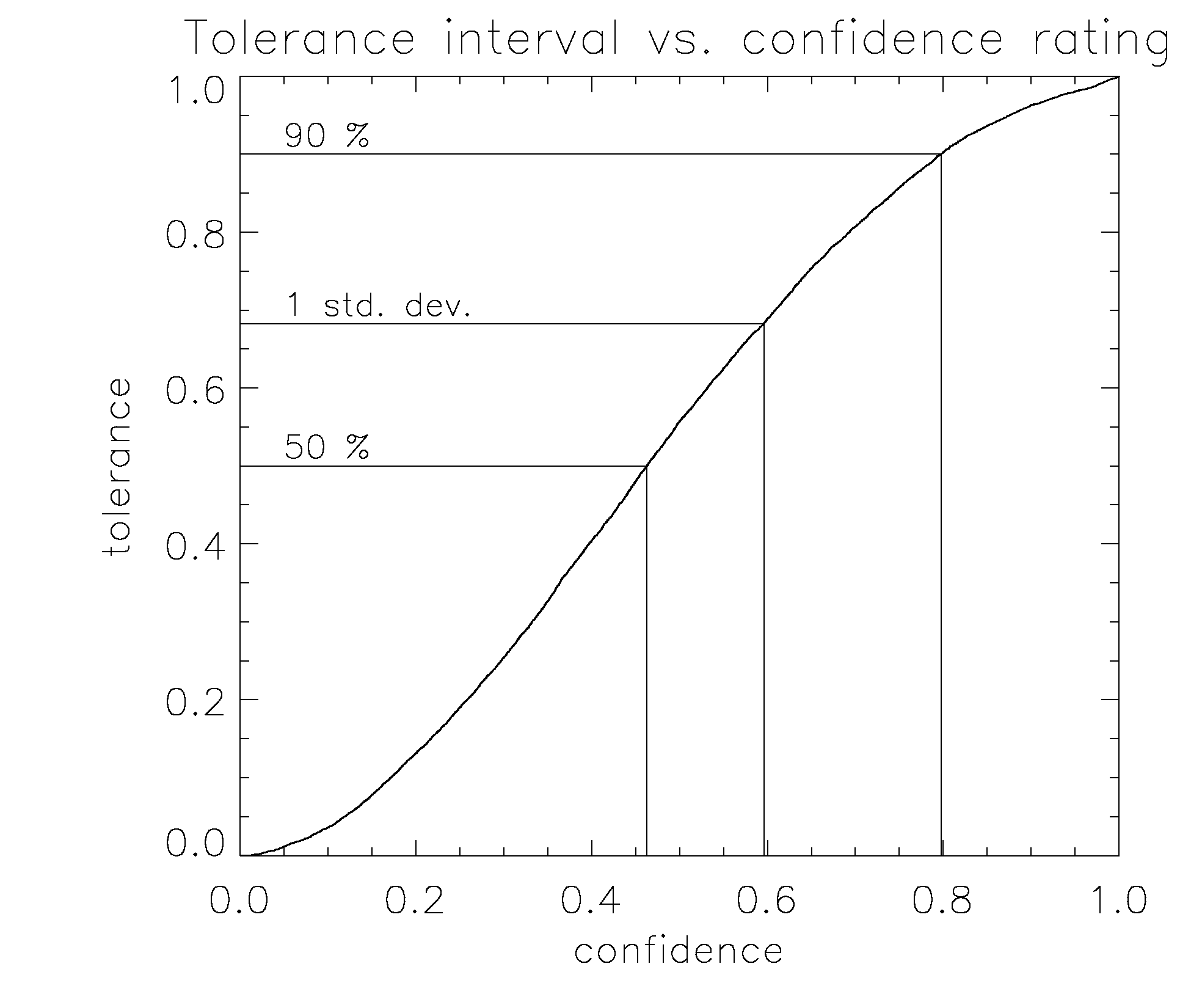}
  \caption{Plot of isoline confidence interval as a function of confidence
  rating.}\label{con_func_plot}
\end{figure}

For performing classifications of water-vapour mixing-ratio for the purpose of
isoline retrieval, the vector $\mathbf{x}$ was defined as follows:
\begin{equation}
\mathbf{x} = \left ( \frac{T_6}{\sigma_6}, 
		\frac{T_7}{\sigma_7}, \frac{T_8}{\sigma_8},
                \frac{T_9}{\sigma_9}, \frac{T_{18}}{\sigma_{18}},
                \frac{T_{19}}{\sigma_{19}}, \frac{T_{20}}{\sigma_{20}}
                \right ) \label{xdef}
\end{equation}
where $T_i$ is the brightness temperature of the
$i$th AMSU channel and $\sigma_i$ is its corresponding standard deviation.  The
first four coordinates supply temperature information, while the last three
supply the humidity information. 
All seven have weighting functions--defined as the gradient of the
brightness temperature with respect to changes in water-vapour
at each vertical level--that
peak in the troposphere with relatively little surface contribution 
on average, especially in the lower latitudes where the air is moist.

The classes are defined as follows:
\begin{equation}
c = \left \lbrace \begin{array}{lc} 1; & q < q_0 \\ 
		2; & q \geq q_0\end{array} \right .
\end{equation}
where $c$ is the class, $q$ is mass-mixing-ratio of water vapour 
(specific humidity) and $q_0$ is the isoline we are trying
to retrieve.

To supply the training data, 86 000 profiles were randomly
selected in both time and space from the ECMWF data.  Radiative transfer
simulations were performed using RTTOV, including both scattering and clouds, to
generate the brightness temperatures needed in (\ref{xdef}).  For ice
cloud simulations, we used aggregrate crystal shapes having a size 
distribution as described in \citet{Wyser1997}.  This
training data will also be used for the actual retrievals.

The retrieval was performed along ECWMF sigma level \#36 (approx. 400 hPa)
 since this is roughly in the centre of
the region of sensitivity for the seven channels.  
Retrievals performed on isentropic surfaces are not accurate because
altitude tends to increase with latitude while the instrument 
weighting functions do the opposite.
To determine the threshold
value for water-vapour, we first look at the statistics, as shown in Figure
\ref{wv_vmr_hist}.  Note the extended tail, with the discontinuity 
($q_0=0.001$ mass-mixing-ratio) being the best threshold.
Selecting the threshold in the valley between the peaks 
of a bi-modal (or in this case, almost bi-modal) distribution
should reduce the error rate because the 
continuum value is more likely to occur near 
one of the two peaks, away from the threshold.

Once classifications have been done over a large enough section of the
Earth's surface, isolines are retrieved by simply tracing the borders
between positive (higher than the threshold) and negative (lower than
the threshold) classification results via any contouring algorithm.  
The results of a simulated test retrieval are shown
in Figure \ref{sample_ret1} using the aforementioned training
data.  To compute the test data, additional radiances
were calculated from a global ECMWF field 
in the same manner as for the training data.
To account for differences in surface emissivity, the test data was
divided into land and sea and retrieved using two, separately simulated
sets of training data.

The heavy black line in Figure \ref{sample_ret1}
is the ``true'' isoline, with auxiliary fine contours for higher and
lower multiples.  The retrieved isoline is not shown;  rather a dot
indicates a positive classification.  Obviously, all the dots should fall
within the heavy contour, but like a young child who has not yet learned
how to colour, the retrieval does not perfectly fill the isoline.

The shading indicates the confidence rating which we define by simply
rescaling the conditional probability:
\begin{equation}
C = \frac{n_c P(c | \mathbf{x}) - 1}{n_c -1} = | R |
\end{equation}
where $c$ is the winning class and $n_c=2$ is the number of classes.  If C
is zero, then the classification result is little better than chance,
while if one then it should be perfect, assuming that the
conditional probability has been estimated accurately.
\begin{figure}
  \centering
  \includegraphics[angle=90, width=0.95\textwidth]{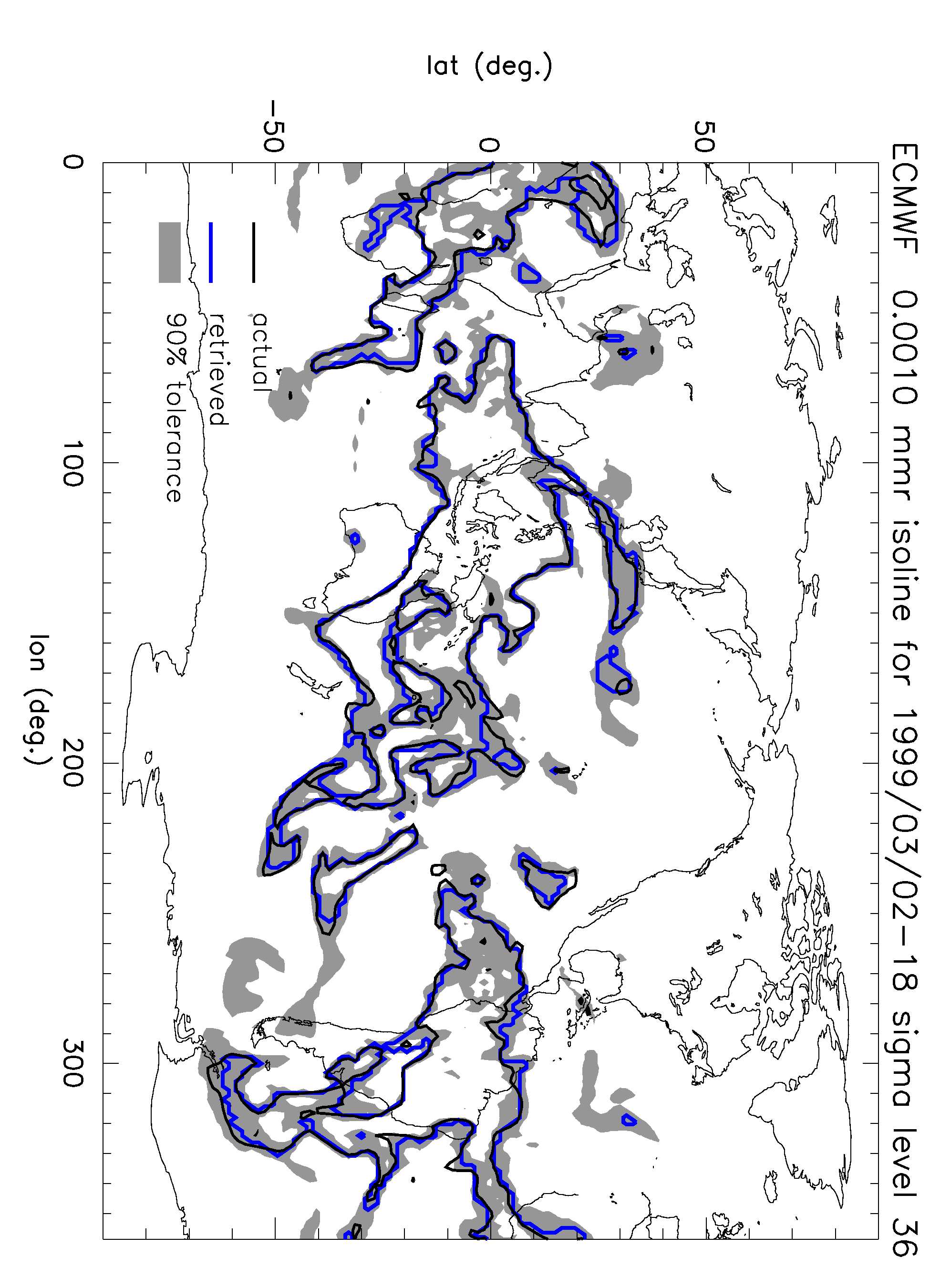}
  \caption{Simulated isoline retrieval, showing tolerance interval.}\label{sample_ret_con}
\end{figure}
As might be expected, the confidence is lower closer to the isoline, as
well as where the gradients are shallow.  The measure may be used to 
define an interval within which the true isoline is
likely to fall:  a ``cutoff'' value is chosen and the area thus enclosed
will contain the true isoline to within a
certain statistical tolerance.

To calculate the tolerance from the confidence rating, a rather
tedious line-integral method was used, expressed mathematically 
as follows:
\begin{equation}
\delta_l(C) = \frac{1}{l} \int_0^l \mathrm H \left [C - C^\prime(\mathbf{r}) \right ] ds
\label{con_func}
\end{equation}
where $\mathrm H$ is the Heaviside function, $\delta_l$ is the tolerance 
of the retrieved isoline as a function of confidence rating and
$C^\prime$ is the confidence rating of the classification results
as a function of geographical position, $\mathbf{r}$.  The integral is
performed along the length, $l$, of the actual isoline.  Although
(\ref{con_func}) implies that the integral must be evaluated separately
for each value of the confidence rating, $C$, in actual fact it may be
done for all values of $C$ by sorting the confidence ratings of the
results, i.e., $C^\prime$.

Figure \ref{con_func_plot} plots the combined results of Equation (\ref{con_func}) for
six (6) different simulated retrievals.  Fifty percent, ninety percent
and one-standard deviation (assuming the statistics are Gaussian) intervals
are indicated in the figure.  This result will be used for the actual
retrievals and in Figure \ref{sample_ret_con} which, while containing
 nothing that Figure \ref{sample_ret1} does not, demonstrates how to set a 
definite tolerance limit on the retrieved isoline.  Now 
the solid line represents the true isoline 
which we expect to fall within the shading ninety percent of the time,
while the broken line is the retrieval.

\section{Results}

Since the AMSU instrument views the Earth at many different
angles to produce a swath, several sets of training data 
corresponding to some or all of these
 viewing angles must be employed to perform actual retrievals.
Radiative transfer simulations were performed at sixteen different
angles, one at nadir and one for every third angle of the `B' 
instrument along one side of the track.  Results are interpolated
for intermediate angles.

Retrievals were performed at the resolution of AMSU-B by interpolating
the AMSU-A data to its equivalent.  Since the `A' instrument detects
temperature which has a diffusion mechanism
(radiative transfer) not available to other tracers, we
expect the small-scale variations to be less pronounced.  Therefore,
the final resolution should be closer to that of the `B' instrument.

\subsection{Interpolation}

Before generating the contours for both the isoline and the confidence interval,
evenly gridded values must be produced at fixed times.  Since the
satellite data is not as irregular as it first seems, interpolates can be
calculated using a simple modification of the multi-linear method.
Scan tracks generate samples along a grid that is rectangular to good
approximation, therefore we search both forward and backward in time
for the two nearest swaths having at least four samples that spatially
enclose the interpolation point.
Bilinear interpolation is performed for both times and the 
final result linearly interpolated between these two values.

A more accurate result could be obtained by accounting for the motion
of the air parcel, but since we are trying to validate a simulation 
based on wind circulation, this would obviously contaminate the results.

Conditional probabilities have been derived for a specific humidity 
(mass-mixing-ratio) threshold of 0.001 for each measurement pixel over 
a nine-day period from 1 September 2002 to 9 September 2002.  
These in turn were interpolated 
to a rectangular lon-lat grid with a resolution of 0.2 degrees 
at twelve-hour intervals 
over a period of five days 
starting at the third of September.  
The two-day overshoot in the initial time interval
is necessary for the interpolation procedure.  The maximum interval between
adjacent swaths ahead of and behind the interpolation times
was slightly over two days, while the average was roughly six hours.

The results for the 12 hour retrievals are shown in Figures \ref{results1}
through \ref{results5}.  The top of each plot displays the retrieval versus the
ECMWF isoline with the grey shading enclosing the 90\% tolerance while 
the bottom shows the advected contours with the differential shading 
indicating the confidence rating.  Contours were advected
using 4 Runge-Kutta steps every six hours after which the they
were re-interpolated so that each pair of points traced out a minimum
fraction of arc of 0.5 degrees and a maximum of one degree.
Contours were initialised with isoline retrievals from the 3 September,
00 UTH.

\subsection{Calibration and validation}

\begin{figure}
  \centering
  \includegraphics[angle=90, width=0.95\textwidth]{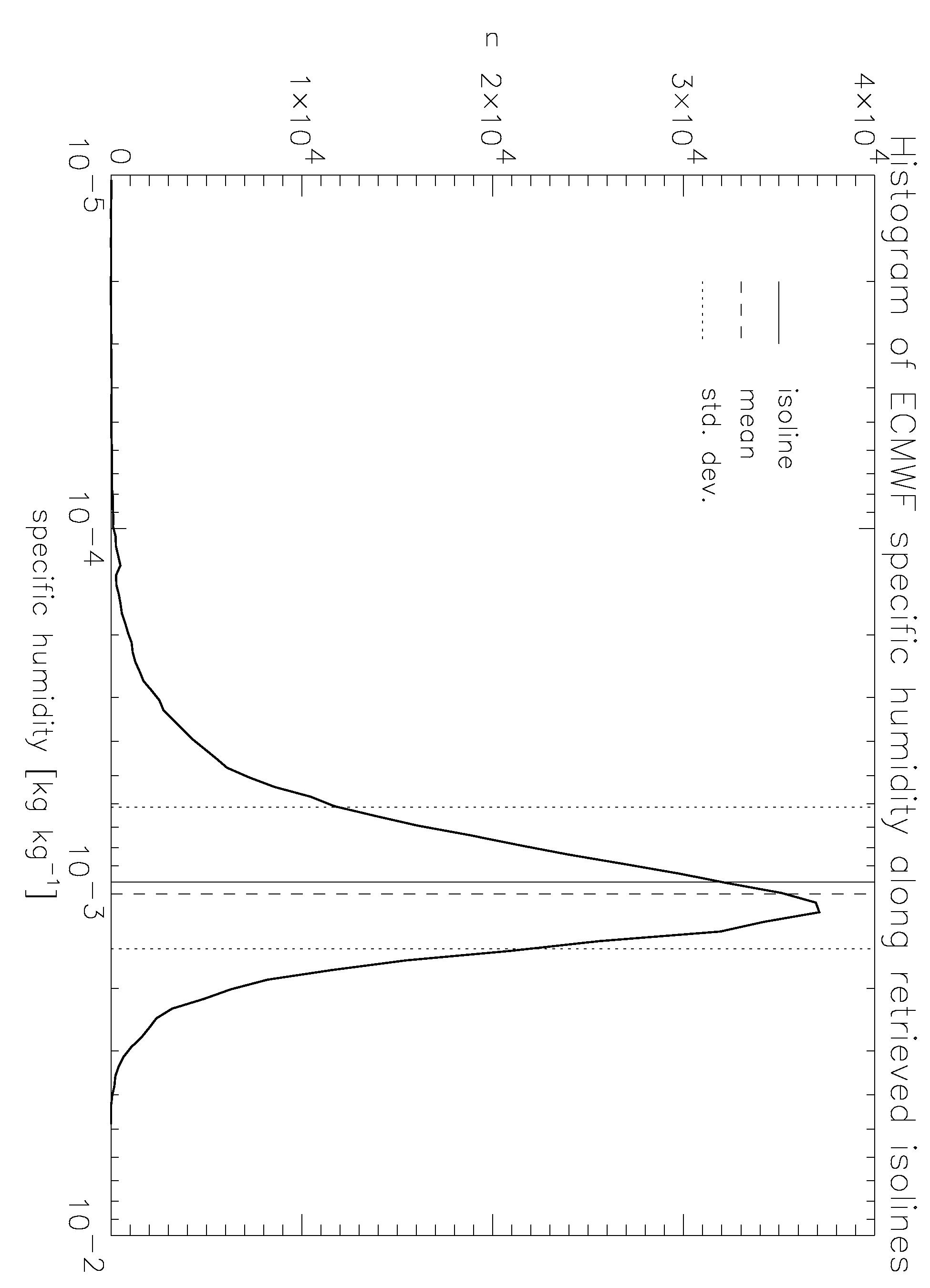}
  \caption{Histogram of ECMWF water-vapour mixing-ratios along retrieved isolines.}\label{ret_hist}
\end{figure}

The distribution of ECMWF specific humidity values along the retrieved isolines
is shown in Figure \ref{ret_hist}.  The bias is $8.1 \times 10^{-5}$ while the 
standard deviation is $4.7 \times 10^{-4}$.

\begin{figure}
  \centering
  \includegraphics[width=0.95\textwidth]{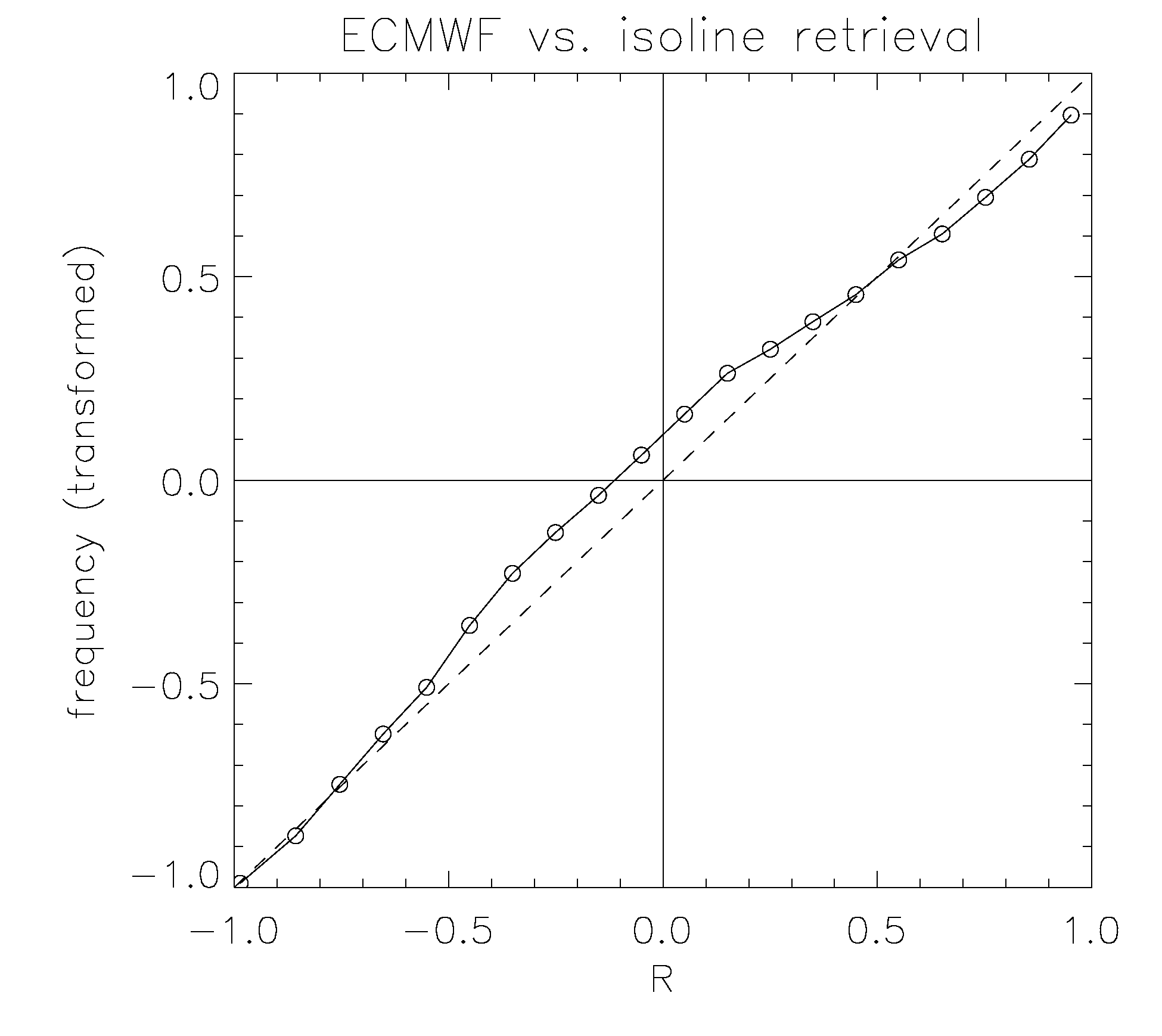}
  \caption{Frequency of ECMWF-derived class data plotted against conditional probability.}\label{accuracy_ecmwf}
\end{figure}

Retrieved conditional probabilities were also validated against
the ECMWF data by comparing them with the
class frequency.  The conditional probabilities were binned at even
intervals and an average taken.  The relative accuracy or frequency 
for each bin was then
computed for the corresponding ECMWF-derived classifications and plotted with
respect to the former as shown in Figure \ref{accuracy_ecmwf}.

\begin{table}
\caption{Effect on classification accuracy of moving the class border.}
\label{acc_rescale}
\centering
\begin{tabular}{|c|c|c|c|} \hline
threshold & \multicolumn{3}{|c|}{accuracy} \\ \hline
value of R & class 1 & class 2 & overall \\ \hline
-1.0 & 0.001 & 1.00 & 0.238 \\
-0.9 & 0.821 & 0.976 & 0.844 \\
-0.8 & 0.879 & 0.953 & 0.890 \\
-0.7 & 0.907 & 0.930 & 0.910 \\
-0.6 & 0.925 & 0.906 & 0.922 \\
-0.5 & 0.937 & 0.882 & 0.929 \\
-0.4 & 0.947 & 0.856 & 0.933 \\
-0.3 & 0.954 & 0.828 & 0.936 \\
-0.2 & 0.961 & 0.800 & 0.937 \\
-0.1 & 0.966 & 0.771 & 0.937 \\
 0.0 & 0.971 & 0.741 & 0.937 \\
 0.1 & 0.975 & 0.709 & 0.936 \\
 0.2 & 0.978 & 0.675 & 0.933 \\
 0.3 & 0.981 & 0.638 & 0.931 \\
 0.4 & 0.984 & 0.597 & 0.927 \\
 0.5 & 0.987 & 0.552 & 0.923 \\
 0.6 & 0.990 & 0.498 & 0.918 \\
 0.7 & 0.993 & 0.434 & 0.916 \\
 0.8 & 0.995 & 0.350 & 0.900 \\
 0.9 & 0.998 & 0.225 & 0.884 \\
 1.0 & 1.00  & 0.000 & 0.853 \\ \hline
\end{tabular}
\end{table}

When considered strictly in terms of the conditional probabilities, it becomes
possible to make the retrievals essentially perfect by re-scaling the
probabilities to match actual frequencies. \citep{Jolliffe_Stephenson2003}
 Visual inspection of the retrievals
suggests a dry bias in comparison to the ECMWF, as the retrieved isolines
usually fall inside (on the high humidity side) of the ECMWF contours.  
This is confirmed in the
plot as the actual frequency is higher at the probability 
corresponding to the isoline ($R=0$).  Table \ref{acc_rescale} shows
the effect on classification accuracy relative to the ECMWF
of rescaling the value of $R$ at $R=0$, i.e. at the class border.

\begin{figure}
  \centering
  \includegraphics[angle=90, width=0.95\textwidth]{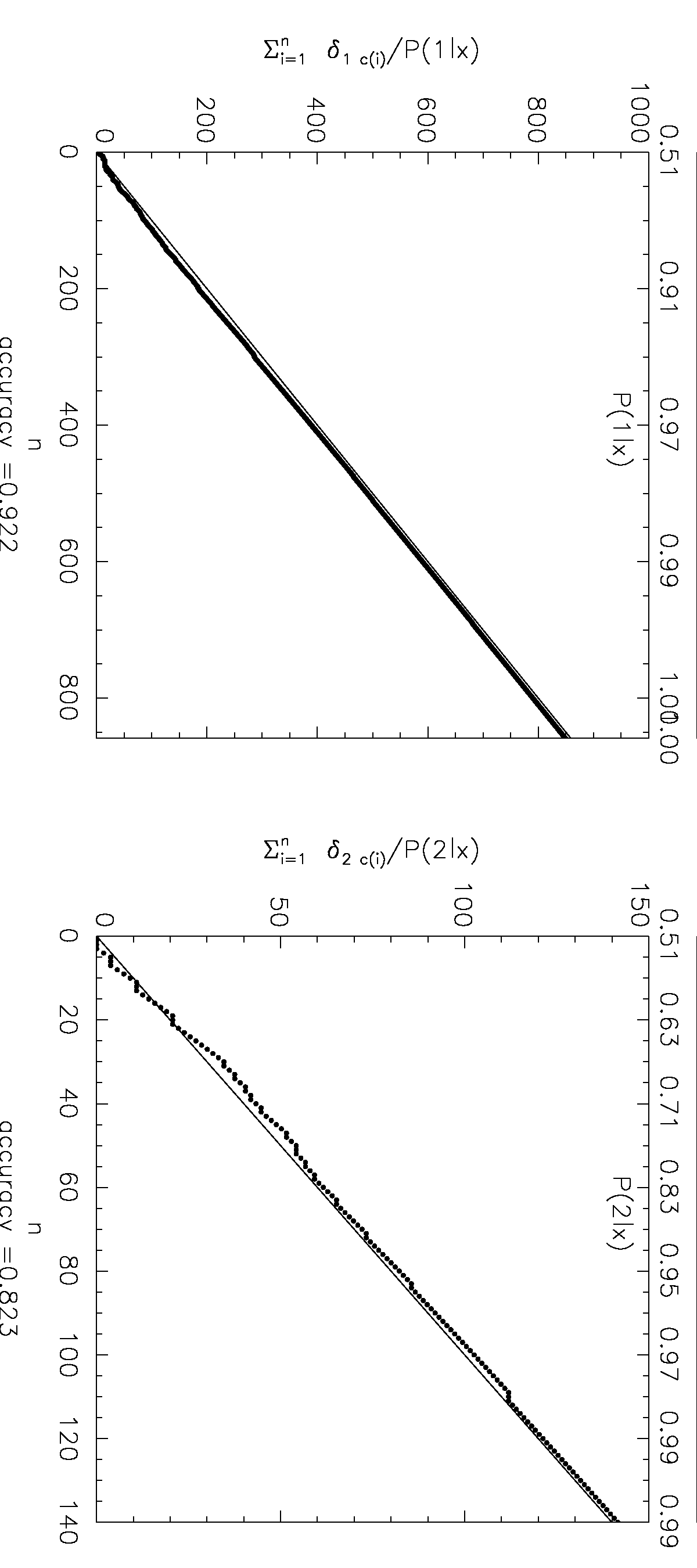}
  \caption{Accuracy of conditional probabilities as compared to radiosonde measurements.  
Cumulative sum of class (as a `yes' or `no' value) divided by conditional probability should follow, on average, one-unit intervals.  Left plot is for retrieved class values of 1, right for 2.}\label{isoret_vs_radiosonde}
\end{figure}

The conditional probabilities were similarly validated relative to radiosonde measurements.
Since there were not enough measurement points to generate reliable statistics
by binning, an integral method was used.  From Bayes' theorem:
\begin{equation}
\int P(\mathbf{x})P(j|\mathbf{x})d\mathbf{x}=\lim_{n \rightarrow \infty} \frac{n_j}{n}
\end{equation}
where $n_j$ is the number of classes with value $j$.  Transforming this
into a Monte Carlo sum:
\begin{equation}
n_j \approx \sum_i P \left [j|\mathbf x(\mathbf{r_i}) \right ]
\label{mc_cp_val}
\end{equation}
where $\mathbf x(\mathbf{r_i})$ is the satellite
measurement vector at measurement point $\mathbf{r_i}$.
The relationship should hold regardless of the particular choice of
points.  By sorting the probabilities and then comparing
their cumulative sum with the number of classes, we gain an idea
of the relative bias at each value.

This will generate a distorted graph, however:  points at the beginning 
of the trace will be much more closely
spaced than those toward the end.
Therefore, we rearrange Equation (\ref{mc_cp_val}) as follows:
\begin{equation}
n \approx \sum_{i=1}^n \frac{\delta_{jc(\mathbf{r_i})}}
		{P \left [ j|\mathbf x(\mathbf{r_i}) \right ]} 
\end{equation}
where $\delta$ is the Kronecker delta and $c(\mathbf{r_i})$ is the
``true'' class (derived from radiosonde measurements) at point $\mathbf {r_i}$.  
Now the cumulative sum will follow, on average,
one-step intervals between each point in the analysis.  To prevent
singularities at low values of the conditional probabilities, the 
analysis was done separately for each class in the retrievals, 
as shown in Figure \ref{isoret_vs_radiosonde}.  
1001 launches from 215 stations--primarily in the Northern Hemisphere--were 
used in the analyis.
The plots deviate very little from
the diagonal, implying that there is little bias relative 
to the radiosonde measurements.  Overall classification
accuracy was 90.8\%.  
The location error in the radiosonde measurements is one factor limiting
the accuracy of these comparisons since balloon drift has not been
accounted for.

\begin{figure}
  \centering
  \includegraphics[width=0.95\textwidth]{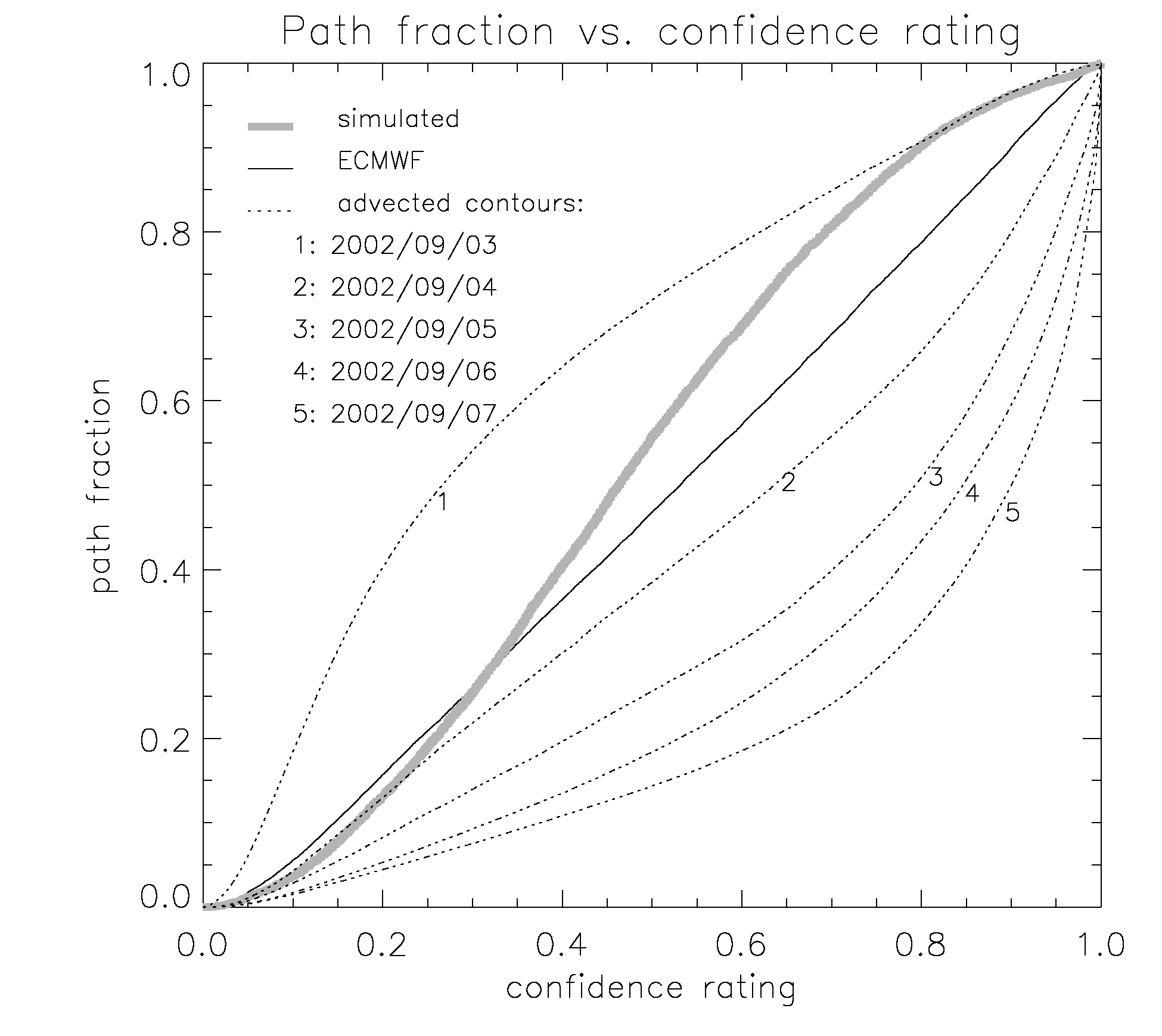}
  \caption{Confidence rating related to fraction of path for ECMWF and advected contours, as compared to simulated retrievals.}\label{conintfunc3}
\end{figure}

Finally, the retrievals were validated by applying the line integral
in (\ref{con_func}) to both the ECMWF isolines and to the advected
contours using the retrieved confidence ratings.  
These are compared in Figure \ref{conintfunc3} 
to the same function, first presented in Figure \ref{con_func_plot},
derived from simulated retrievals.  While the
ECMWF isolines were integrated for the entire span of five days, 
the advected contours were integrated day by day for both the 0 hr
and 12 hr retrievals to see the change in accuracy over time.

The confidence rating will approach zero along the isoline,
thus the more closely the reference isoline matches the retrieved, 
the further up and to the left 
the function will be, with an exact match returning the Heaviside
function.  The thick, grey line from the simulation
runs provides an upper limit to
retrieval accuracy:  any new sources of error, either in the retrieval
or in the reference, will move the integral down and to the right.
Since advected contours were initialised with retrieved isolines, the
very first ones lie to the left.

\section{Discussion}

\begin{figure}
  \begin{center}
  \includegraphics[width=1\textwidth]{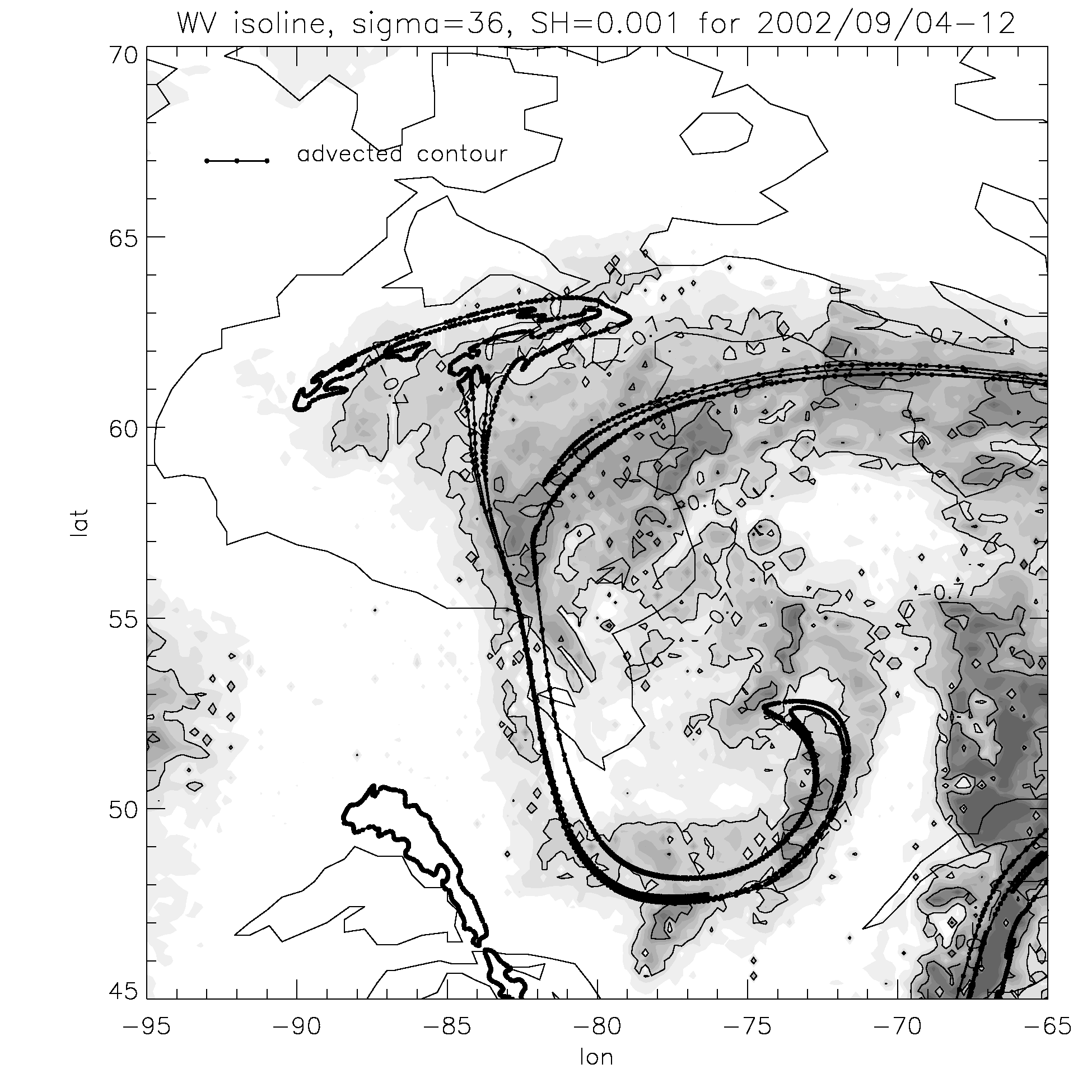}
  \includegraphics[width=0.8\textwidth]{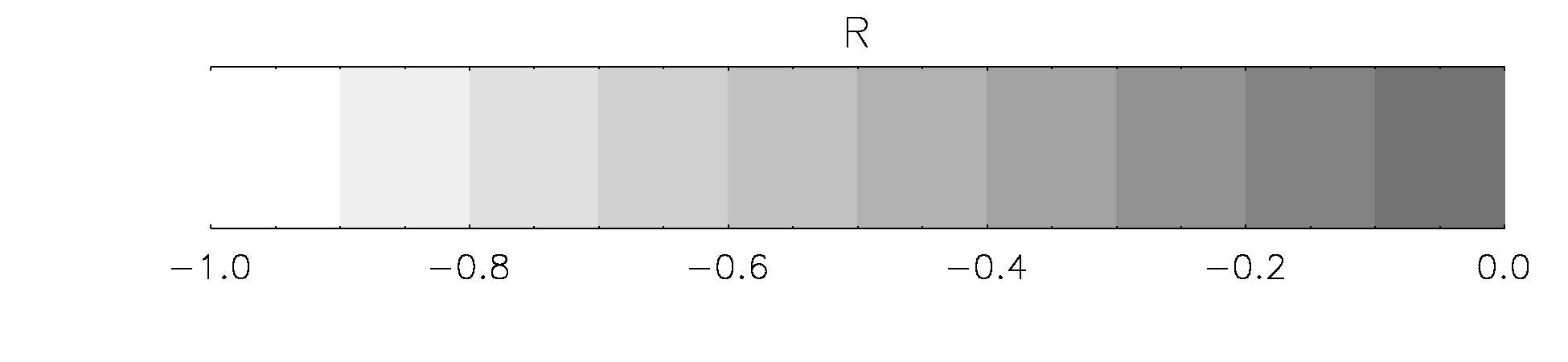}
  \end{center}
  \caption{Detail of 4 September retrieval with advected contour.}
  \label{detail}
\end{figure}

The study of chaotic mixing is interesting, not only in its own right,
but also for its potential to improve fluid modelling in geophysical
systems.  It is particularly relevant to the parameterisation of 
large-scale diffusion in circulation models, since mechanical 
stirring enhances diffusion. \citep{Nakamura2001, Thiffeault2003}
A unique approach to this phenomenon
is the modified Lagrangian mean formalism by \citet{Nakamura1998}.
A set of retrieved contours would lend themselves well to this
type of analysis.

But first we must answer the question addressed by this paper:
is there evidence of chaotic mixing in the satellite data?
Looking strictly at the isoline retrievals proper, the answer is,
``no.''  Evidence of fine-scale mixing is, however, apparent in the
conditional probabilities.  Many of the filaments seen in the 
advected contours also show up as traces of reduced ``confidence
rating.''  In other words, there is a non-negligible probability
(less than 50 \%) that the filament might be there.
Recalling how the retrieval may be recalibrated: by changing
the threshold value of R at which the isoline is drawn, 
the filaments are thereby exposed with relatively little effect
on the larger ``islands.''  See for example the expanded detail 
of the anticyclone in the North-West corner of the 4 September retrieval
(Figure \ref{results2}) in Figure \ref{detail}.

There are a number of difficulties and sources of error associated
with this study.  The advected contours especially are of limited
accuracy, even over short time scales and they diverge rapidly.
Contours are advected on sigma levels and do not take into account 
non-adiabatic vertical motion, which can be strong in the troposphere.  
The choice of sigma levels over isentropic levels had more to do with retrieval
accuracy but further reduces the effectiveness of the transport model, since
even adiabatic motion in the vertical is now neglected.
Another limitation of contour advection is that it cannot easily
account for sources or sinks which are numerous for water-vapour.
The high value chosen, while making detection easier, will bring with
it a high occurrence of clouds and precipitiation.
There is a further sense in which water-vapour is not a ``passive''
tracer: it will affect the motion.
Repeating the analysis with a similar instrument that detects
ozone in the stratosphere, such as GOME,
would likely produce more satisfying results.

The retrieval method suffers from similar limitations.  In particular,
it attempts to limit the retrieval to a single vertical level,
a feat that is difficult to accomplish with only seven channels to work
with.  This is the main reason why the retrievals are seen to have
such a broad tolerance interval.  Another difficulty is the scattering
generated by clouds, which for high, thick cover may interfere with
the view of the target level.  Adding the two lowest frequency 
AMSU-B channels might help address both issues, but this requires
accurate knowledge of the surface emissivities.

Finally, some minor artifacts are produced by the interpolation procedure
at junctions between scan lines.  These could be cleaned up by some
sort of image processing technique, such as filtering.  Interpolation
could also be done at narrow time intervals and additional filtering applied
in the time domain so as to include extra measurement pixels in each 
grid point.

These and other difficulties only serve to underline the considerable
potential of the inversion method.  Producing continuum retrievals, 
for instance, would be a trivial extension accomplished
by generating several contours and interpolating between 
them.  Since satellite retrievals
are rarely better than 5 \% accurate, the error produced by interpolation
would be negligible in comparison.  It would follow that generating
continuum values from a two-dimensional array of discrete ones would
be more effective than relying solely on individual pixels.
An accurate characterisation of the error at each point in the
field is easily generated as a by-product from the combination
of confidence ratings and tracer gradients. 

\section{Conclusion}

The natural apex of satellite remote sensing is that of
an intimate, real-time picture of the atmosphere
down to the finest detail.
Neural networks, statistical classification and similar machine
learning techniques provide a nice alternative to complex
inverse methods such as optimal estimation that use forward
models directly.  As the models become more detailed and precise,
it becomes increasingly difficult to apply the latter methods with
anything resembling analytical precision.  Machine learning
techniques, by contrast, easily generate an inverse model that
can be applied directly and efficiently to the data.
Although
it is usually impossible to derive analytical inverse functions for complex
radiative transfer models, there is no reason that a numerically
generated one not be used instead.  While the current paper does use
an involved forward model to generate the training data,
as the satellite record becomes longer,
this can be superceded by satellite measurements paired with co-located,
direct measurements, obviating the need for lengthy simulations runs,
while ensuring accuracy.

In this paper we have generated very specific retrievals that detect
from high-resolution AMSU satellite measurements
only a single isoline of water-vapour along a single vertical level
in the upper troposphere.
These retrievals were shown to have good agreement 
both with ECMWF assimilation data and with radiosonde measurements.  
The method is shown to have excellent potential for detecting 
fine-scale features produced by chaotic mixing, although this is
hampered by the poor vertical resolution of the instrument, resulting
in reduced accuracy which in turn decreases the effective horizontal
resolution.


\input{isoline_results.col.tex}

%% file: agf_simple_els.tex
The $k$-nearest neighbours is a popular statistical classification technique
in which a fixed number, $k$, of training samples closest to the test point
 are found and the class determined by voting.  
\citep{Michie_etal1994}
Consider the following generalization of the scheme:
\begin{eqnarray}
P(j | \mathbf{x}) & \approx & \frac{1}{W} \sum_{i, c_i=j} w_i(\sigma) \\
\sum_{i=0}^n w_i(\sigma) &=& W
\end{eqnarray}
where $\mathbf{x}$ is a test point, $W$ is a constant,
$c_i$ is the class associated with the $i$th sample and 
$w_i$ is a weight calculated via a filter function:
\begin{equation}
w_i(\sigma) = g \left (\frac{\mathbf{x}-\mathbf{x_i}}{\sigma} \right )
\end{equation}
where $g$ is the filter function $\sigma$ is its width
and $\mathbf{x_i}$ is the location in measurement space of the
$i$th training sample.

The parameter, $W$, is equivalent to the number of nearest
neighbours in a $k$-nearest-neighbours classifier and is held fixed by
varying the filter width.  This keeps a uniform number of samples
within the central region of the filter.
An obvious choice for $g$ would be a Gaussian: 
\begin{equation}
g(\mathbf{\Delta x})=\exp \left (-\frac{|\mathbf{\Delta x}|^2}{2} \right)
\end{equation}
Where the upright brackets denote a metric, typically Cartesian.

The primary advantage of the above over a $k$-nearest-neighbours, is that 
it generates estimates that are both continuous and differentiable.
Both features may be exploited, first to find the class 
borders, then to perform classifications and estimate the
conditional probability.  Let $R$ be
the difference in conditional probabilities between two classes:
\begin{equation}
R(\mathbf{x}) =  P(2 | \mathbf{x}) - P(1 | \mathbf{x})
\end{equation}
where $1$ and $2$ are the classes.  The border between the two is
found by setting this expression to zero.
The procedure used was
to randomly pick pairs of points that straddle the class border
and then solve along the lines between them.
Analytical derivatives are used as an aid to root-finding.

The class of a test point is estimated as follows:
\begin{eqnarray}
j & = & \arg \underset{i}{\min} | \mathbf{x} - \mathbf{b_i} | \\
p & = & (\mathbf{x} - \mathbf{b_j}) \cdot \nabla_{\mathbf{x}} R (\mathbf{b_j})
\label{peq} \\
c & = & ( 3 + \mathrm{sgn} p) / 2
\end{eqnarray}
where $\{\mathbf{b_i}\}$ sample the class border and $c$ is the retrieved class.
The value of $R$ may be extrapolated to the test point:
\begin{equation}
R \approx \tanh p
\label{confidence_est}
\end{equation}
This algorithm is robust, general and efficient, 
yet still supplies knowledge of the
conditional probabilities needed to set a definite 
confidence limit on the retrieved isoline. It is a variable
bandwidth kernel density ``balloon'' estimator.  \citep{Terrell_Scott1992}

{\color{red}\citet{Mills2011} provides a more thorough description of the AGF algorithm.}

%% file: isoline_results.col.tex
\begin{figure*}
  \begin{center}
  \includegraphics[width=0.95\textwidth]{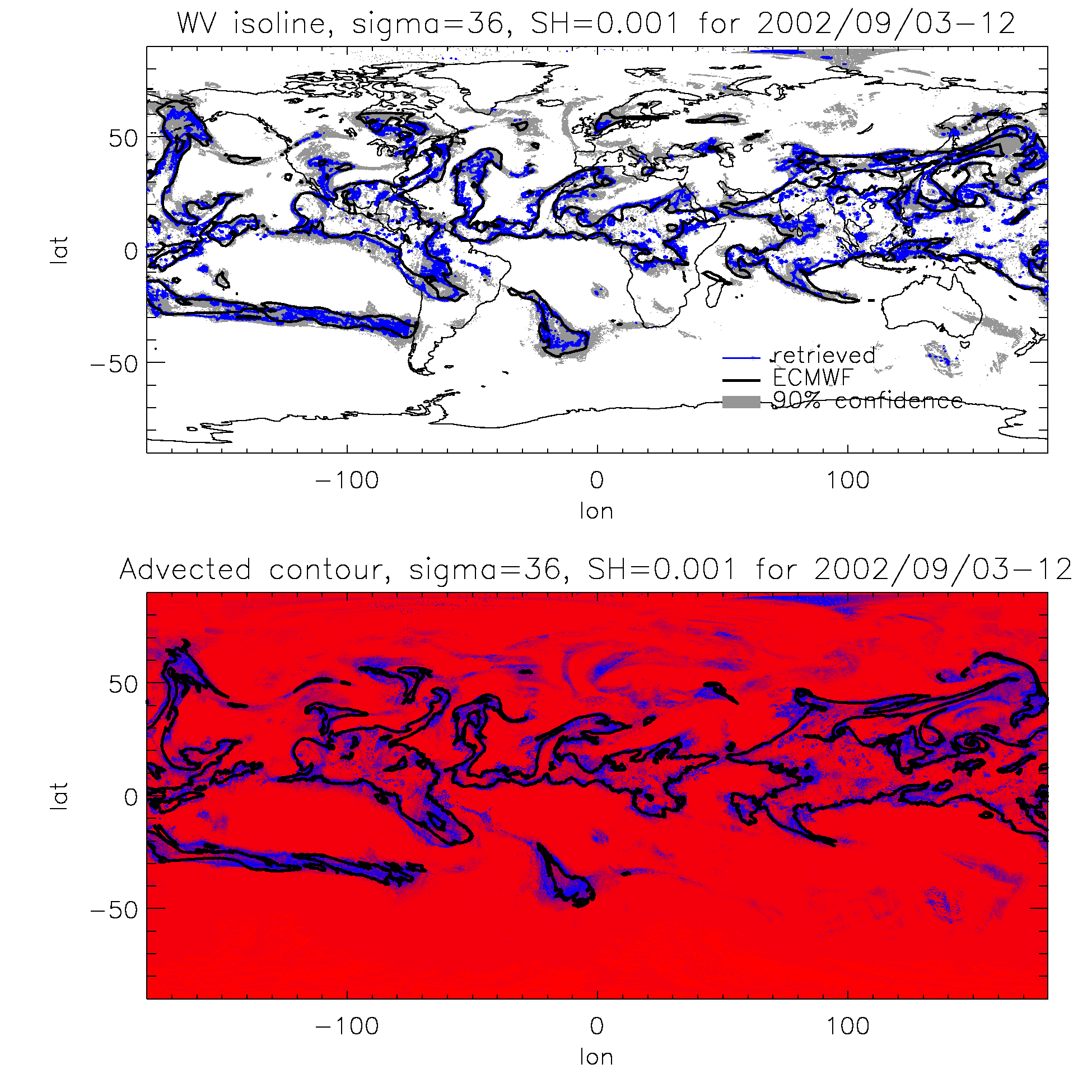}
  \includegraphics[width=0.6\textwidth]{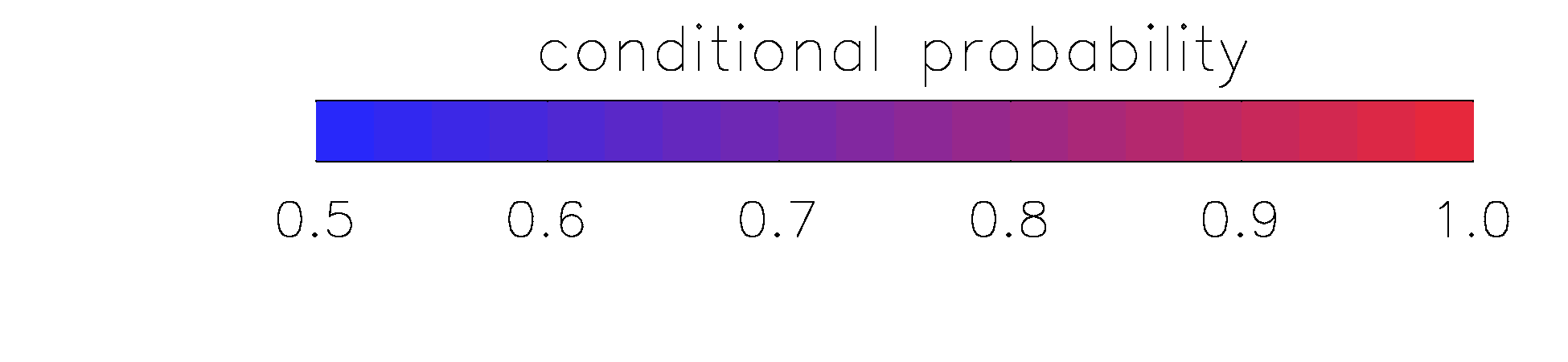}
  \caption{Isoline retrieval (top) vs. contour advection (bottom) 3 Sept 2002 12:00 UTH.}
  \label{results1}
  \end{center}
\end{figure*}
\begin{figure*}
  \begin{center}
  \includegraphics[width=0.95\textwidth]{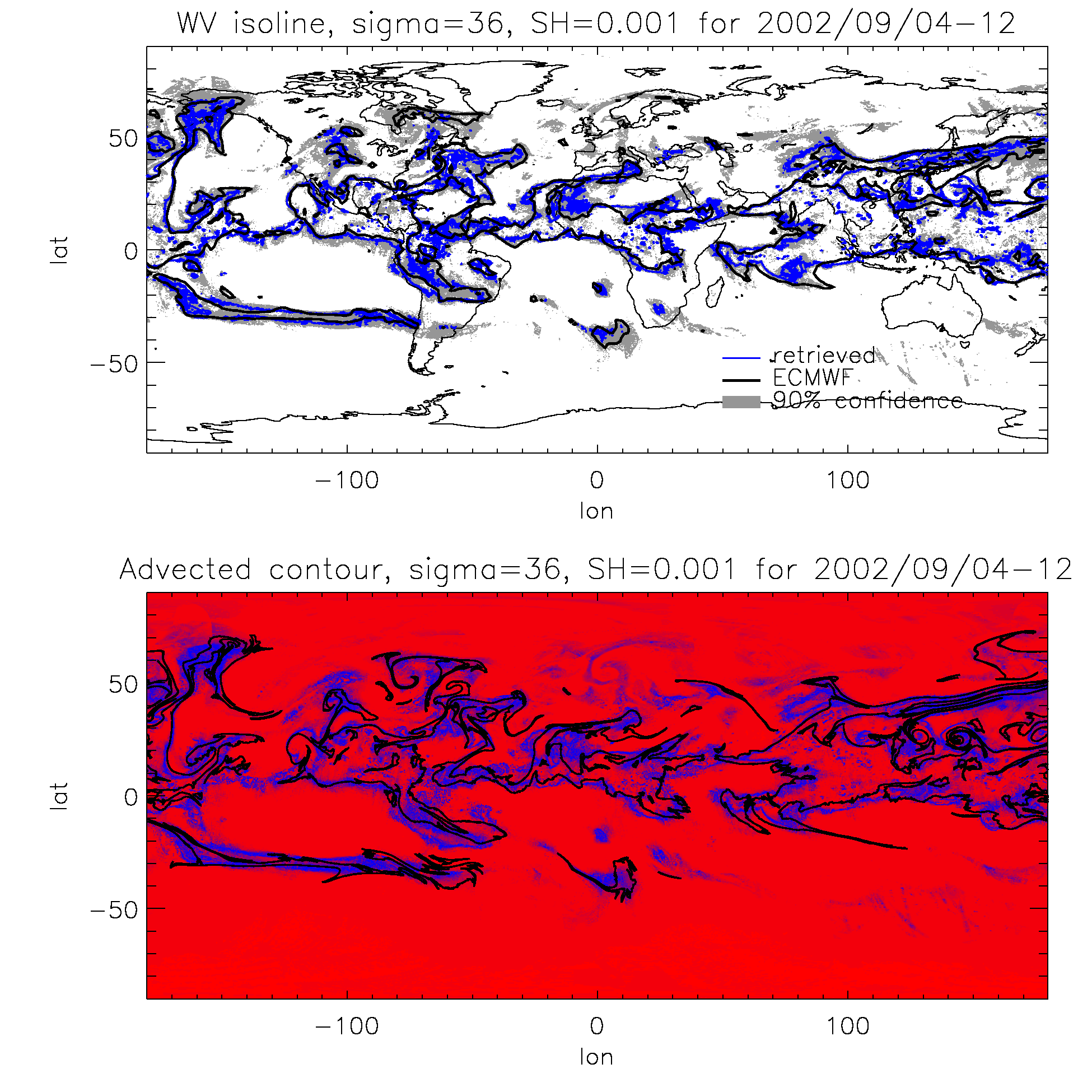}
  \includegraphics[width=0.6\textwidth]{legend_col}
  \caption{Isoline retrieval (top) vs. contour advection (bottom) 4 Sept 2002 12:00 UTH.}
  \label{results2}
  \end{center}
\end{figure*}
\begin{figure*}
  \begin{center}
  \includegraphics[width=0.95\textwidth]{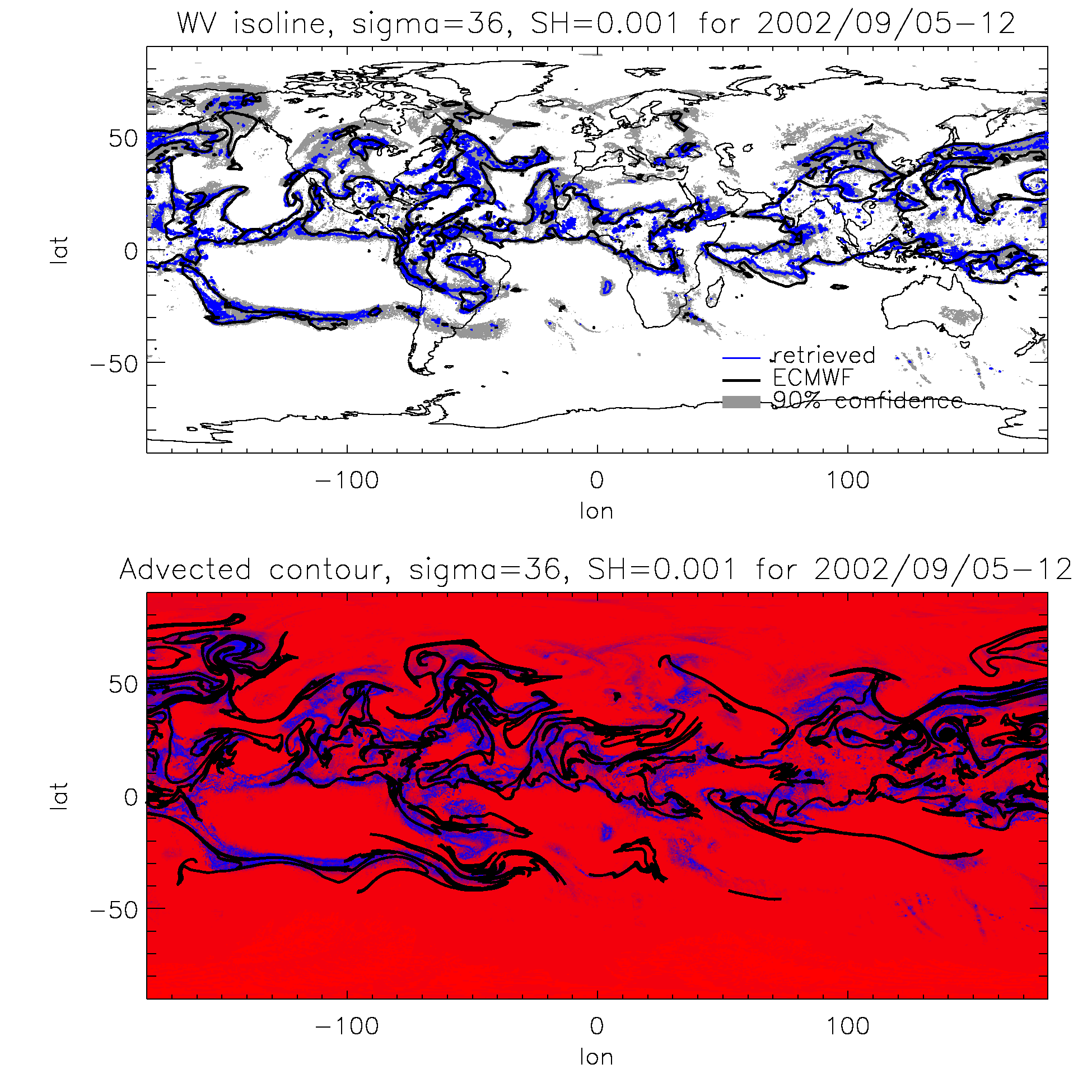}
  \includegraphics[width=0.6\textwidth]{legend_col}
  \caption{Isoline retrieval (top) vs. contour advection (bottom) 5 Sept 2002 12:00 UTH.}
  \label{results3}
  \end{center}
\end{figure*}
\begin{figure*}
  \begin{center}
  \includegraphics[width=0.95\textwidth]{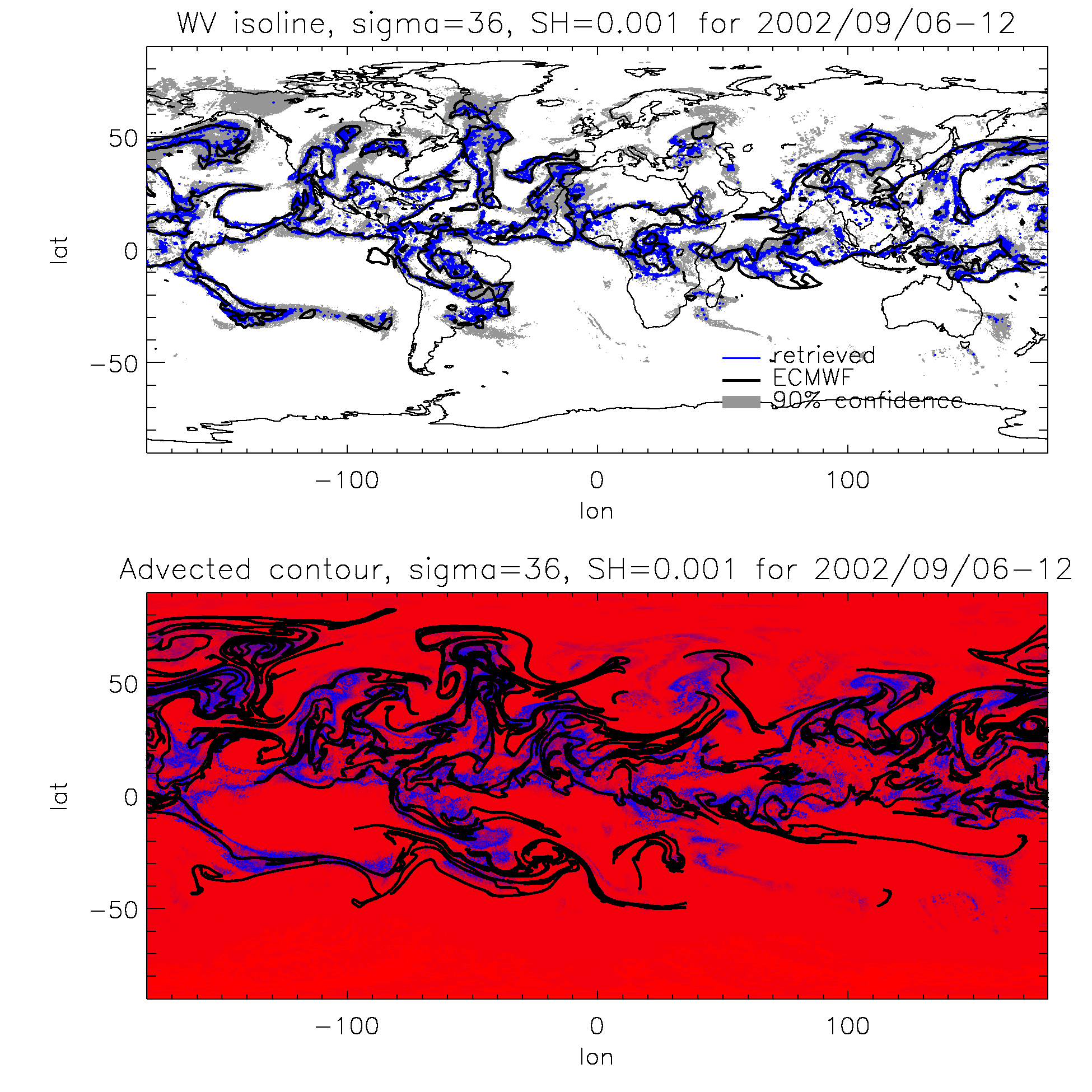}
  \includegraphics[width=0.6\textwidth]{legend_col}
  \caption{Isoline retrieval (top) vs. contour advection (bottom) 6 Sept 2002 12:00 UTH.}
  \label{results4}
  \end{center}
\end{figure*}
\begin{figure*}
  \begin{center}
  \includegraphics[width=0.95\textwidth]{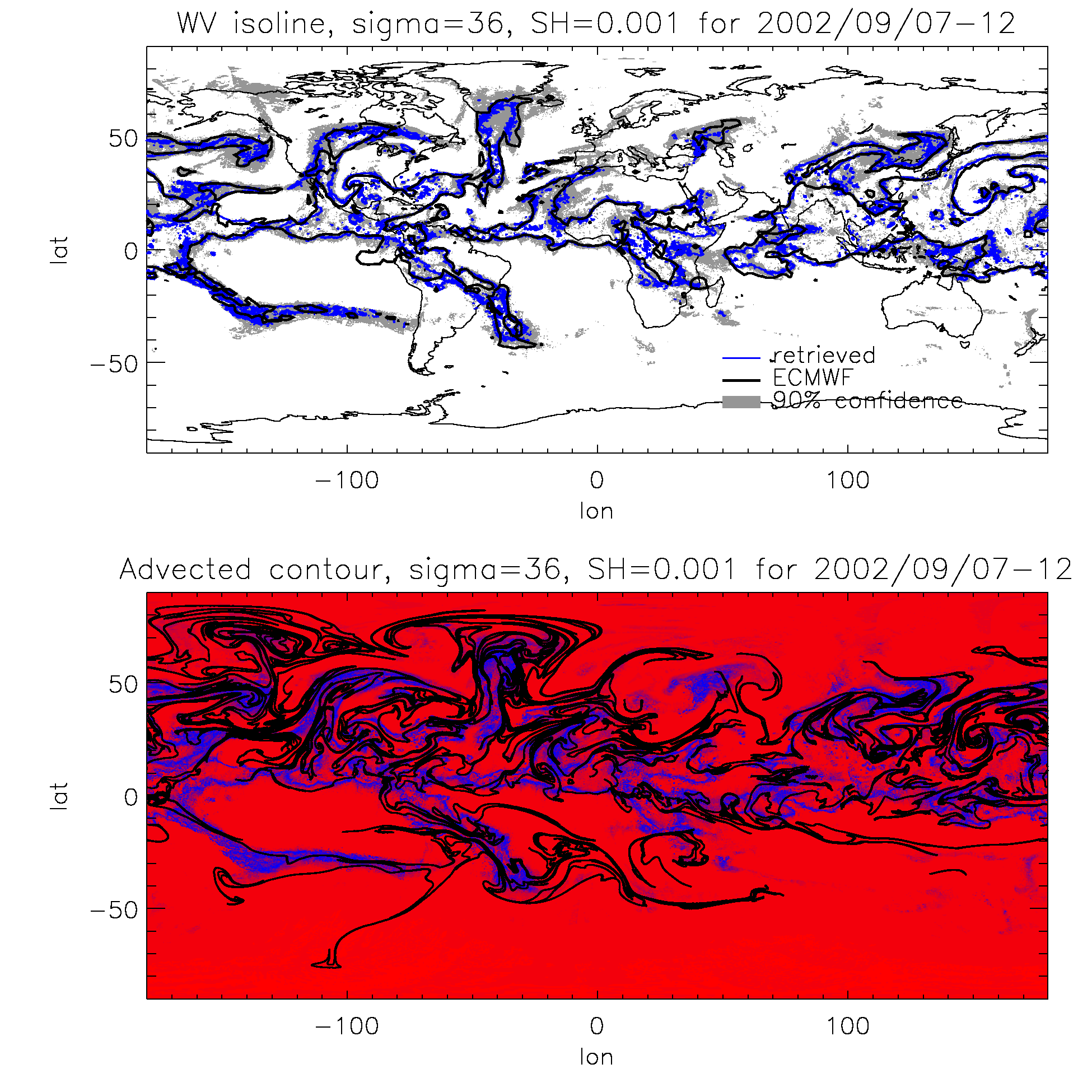}
  \includegraphics[width=0.6\textwidth]{legend_col}
  \caption{Isoline retrieval (top) vs. contour advection (bottom) 7 Sept 2002 12:00 UTH.}
  \label{results5}
  \end{center}
\end{figure*}

%% file: isoline_ack.tex
Thanks very much to my colleagues and supervisors (both present and former)
from the IUP for encouraging and supporting this work, Georg Heygster
and Stefan Buehler in particular.  Thanks to Stefan,
Mashrab Kuvatov and Christian Melsheimer for very valued comments on the
draft manuscript.  

Thanks to ECMWF for ERA 40 data, to the RTTOV development team at the
UK met office, to the Danish Meteorological Institute for the database
of compiled radio soundings and to  Lisa Neclos from the
Comprehensive Large Array-data Stewardship System (CLASS) of the US
National Oceanic and Atmospheric Administration (NOAA) for AMSU
data.

%% file: iso_appendix1.tex
Suppose we have a conditional probability, $P(q | \mathbf{x})$ describing the
distribution of states, $q$, for a given measurement variable, $\mathbf{x}$.
Discretizing the problem into only two
states, with the threshold given by $q_0$,
the continuum distribution transforms as follows:
\begin{eqnarray}
P(1 | \mathbf{x}) = \int_{0}^{q_0} P(q | \mathbf{x}) dq \\
P(2 | \mathbf{x}) =  \int^{\infty}_{q_0} P(q | \mathbf{x}) dq
\end{eqnarray}
If only these discrete states are required, it is very easy to show that a
classification retrieval will be more accurate than one over a continuum.
The accuracy of a series of classification results over an
area, $A$, is given as follows:
\begin{equation}
a = \frac {1}{A} \int_A P \left[c(\mathbf{r}) | \mathbf{x}(\mathbf{r}) \right]
                d\mathbf{r}
\end{equation}
where $c$ is the class as an integer function of position.  Full knowledge
of the conditional probability is assumed.  The value of this integral takes on
a maximum when the integrand is maximised at each point:
\begin{equation}
\max(a) = \frac{1}{A} \int_A \left \lbrace \underset{j}{\max} P \left [j |
        \mathbf{x}(\mathbf{r}) \right ] \right \rbrace d\mathbf{r}
\end{equation}
Since this is simply the definition of maximum likelihood, it is
apparent that any other method of selecting the class will produce 
less accurate results.

Continuum retrievals are generally performed by taking an expectation value:
\begin{equation}
\bar q = \int^{\infty}_{-\infty} q P(q | \mathbf{x}) dq
\end{equation}
where $q$ is the state variable and $\bar q$ the retrieved state variable.
In the continuum case, the class, $c$, is selected quite as follows:
\begin{equation}
  c = \left \lbrace \begin{array}{ll}1;\; \bar q < q_0\\2;\; \bar q \ge
q_0\end{array}\right.
\end{equation}
This will produce different results than a maximum
likelihood classification.